# Hydrostatic pressure to trigger and assist magnetic transitions: baromagnetic refrigeration

M. Quintero[1,2], G. Garbarino[3], A.G.Leyva[1,2]

1-Departamento de Física de la Materia Condensada, GIyA, GAIANN, Comisión Nacional de Energía Atómica, Buenos Aires, Argentina.

2-Escuela de Ciencia y Tecnología, Universidad Nacional de General San Martín, Buenos Aires, Argentina.

3- European Synchrotron Radiation Facility, BP 220, 38043 Grenoble Cedex 9, France.

The possible application of the barocaloric effect to produce solid state refrigerators is a topic of interest in the field of applied physics. In this work, we present experimental data about the influence of external pressure on the magnetic properties of a manganite with phase separation. Using the Jahn Teller effect associated with the presence of the charge ordering we were able to follow the transition to the ferromagnetic state induced by pressure. We also demonstrated that external pressure can assist the ferromagnetic state, decreasing the magnetic field necessary to generate the magnetic transition.

Conventional vapor-cycle refrigerators are based on the barocaloric effect, heating and cooling a fluid by the application of an external pressure. The main drawback of this technology is the necessity of a fluid with large temperature change and with good heat transfer properties, simultaneously[1]. These limitations reduce the available compounds to a small number of gases that are harmful to the environment.

The ever growing demand for refrigeration has motivated intense research into the development of new alternatives to replace existing technology. Magnetic refrigeration has appeared as one of the environmental friendly possibilities.

Magnetic refrigeration is based in the magnetocaloric effect[2] [3] and its temperature change is produced by the application of an external magnetic field on a magnetic refrigerant material. The independence between the refrigerant material and the heat exchanger allows the possibility of using nonhazardous fluids such as water or noble gases.

Despite of the great advances observed in the development of prototypes of room temperature magnetic refrigerators [4], the massive use of this technology in the near future remains uncertain. The challenges that must be overcome are related both to the design of the prototypes and to the selection of the magnetic material [5] . From the point of view of the design, the aspect requiring improvement is the characteristic time related to the heat exchange, which acts as a limitation on the minimum duration of a cycle. Another limitation is the necessity of a high magnetic field to induce a temperature change in the magnetic material[6]. As a consequence, the efficiency of these prototypes is still not high enough to allow general deployment of the technology.

The idea of cooling by adiabatic application of pressure has already been proposed [7] and reported for the Pbnm to R3c structural transition observed in $Pr_{1-x}La_xNiO_3$. A few years ago, Mañosa and co-workers [8]reported a large barocaloric effect in a Ni-Mn-In alloy at room temperature, obtaining values that compare with those obtained in giant magnetocaloric materials. This discovery triggered the idea to use the barocaloric effect in solid state refrigeration combined with the application of a fixed magnetic field to generate a magnetic moment that could be controlled by the application of external pressure [9] [10] [11]. The thermodynamic cycles require a strong coupling between the magnetic and structural degrees of freedom to ensure that a large magnetization change is produced when an external pressure is applied.

The number of systems that exhibit a strong coupling between the magnetic and structural degrees of freedom is quite large, but the complexity of a system tends to increase as the efficiency grows. Manganites represent one of the most studied systems because of the rich variety of degrees of freedom coupled between them.

In this work, we present a structural and magnetic characterization of the manganite $La_{0.625-y}Pr_yCa_{0,375}MnO_3$ with a Pr concentration $y = 0.3$. For this particular concentration, the manganite exhibits a coexistence of regions with different magnetic ordering, the so called phase separation phenomena. The coupling between the magnetic and structural degrees of freedom will be exposed and discussed as a function of the applied external pressure. We will also describe the magnetocaloric properties enhancement by the application of external pressure.

Polycristalline samples of $La_{0.325}Pr_{0,3}Ca_{0,375}MnO_3$ were obtained by a liquid mix method that has been described elsewhere [12]. Magnetic measurements were performed in a VSM magnetometer model Versalab manufactured by Quantum Design. Hydrostatic pressure was applied using a high pressure cell manufactured by HMD model CC-SPr-8.5-D-MC4-1.

X ray diffraction (XRD) data at different temperatures and ambient pressure were obtained using a Rigaku diffractometer. High pressure XRD measurements at room and low temperature were conducted at the beamline ID27 of the European Synchrotron Radiation Facility (ESRF). Powder samples were loaded in a large opening membrane-driven diamond anvil cell (DAC) using stainless-steel gaskets and filled with helium as the pressure-transmitting medium. The monochromatic X-ray beam at 33 keV was focused down to 3×3 $\mu m^2$ and a MARCCD 165 detector was used to collect the diffraction data. For the low-temperature experiment, the DAC was installed in a home made continuous He-flow cryostat. The pressure was increased *in situ* at constant T and measured using the fluorescence of a ruby sphere placed next to the sample. A full-pattern Rietveld refinement of the XRD data was done using the gsas package [13]

The compound used in this work belongs to the widely studied system $La_{0.625-y}Pr_yCa_{0,375}MnO_3$ [14], characterized by the presence of a strong phase separated state in the intermediate Pr region.

In the phase separated region, the system is formed by two different phases: a ferromagnetic metallic (FM) and a charge ordered (CO) phase. One of the most interesting aspects to take into account is the behavior of the charge carriers in each of the mentioned phases. In the case of the ferromagnetic phase, the magnetic ordering of the Mn ions favors the delocalization of the carriers, giving place to a metallic character where the charge is distributed uniformly between the Mn ions. In the CO phase the electrons tends to stay localized in the Mn core, giving place to a clear distinction between $Mn^{+3}$ and $Mn^{+4}$. The presence of this extra electron in the $Mn^{+3}$ ion favors the splitting of the $e_g$ and $t_{2g}$ orbitals inducing a deformation of the $MnO_6$ octahedra known as the Jahn Teller (JT) effect. It is well established that this deformation introduces a modification in the crystal structure of the material.

With these considerations in mind, we can now analyze the system behavior with temperature, magnetic field and pressure.

At ambient pressure and low magnetic field the system presents a room temperature paramagnetic insulator (PI) state followed by a CO phase around $T_{CO}$ = 220 K. The evidence of the presence of the CO phase is found in a small shoulder at $T_{CO}$ observed in the magnetization curve (figure 1). At lower temperature, the FM phase appears at $T_1$ around 200 K coexisting with the CO phase. This phase separated state remains down to a temperature between 70 and 100 K (called $T_2$) where the non-FM region becomes FM giving place to a homogeneous state.

The application of an external magnetic field or hydrostatic pressure favors the development of the FM state and suppressing progressively the step increase of the magnetization observed at $T_2$ .This suppression is a clear indication of the strong coupling between structural, magnetic and electronic degrees of freedom. While the magnetic field favors the development of the ferromagnetic state, hindering the localization of carriers and suppressing the JT distortion. The application of pressure obstructs the formation of the JT distortion, suppressing the localization of charges and making the development of the CO state unfavorable.

The presence of the CO phase is also reflected in the temperature dependence of the lattice parameters (figure 2a). Down to 225K, the lattice parameters show a monotonous reduction. On cooling below T = 225 K, this behavior is affected by the JT distortion associated with the CO phase. For $70K<T<225K$, the *c* axis increases, the *b* axis decreases and the *a* axis remain almost unchanged. At 75K, the FM phase is established and the lattice distortions disappear.

It is clear that the application of a magnetic field or external pressure affects the formation and stability of the CO phase. Considering that the thermodynamic cycles used for refrigeration usually require a stable temperature at the beginning of the cycle (the temperature change is produced by the pressure and not by an external temperature controller), we wanted to know whether it would be possible to suppress the CO phase once it has developed in the sample. To answer this question, we studied the pressure evolution of the structure at a fixed temperature of 150 K, in the phase separated region (see figure 2b).

As the pressure is increased we can clearly observe a jump of the lattice parameters at 0.5 GPa. The a and c axis are reduced by ~0.1% and ~0.2%, respectively, while the b axis expands by ~0.05% . At higher pressure, up to 5 GPa, the lattice parameters do not show any other modification.

To highlight the anomaly in the volume compression and to clarify the order of the transition, the cell volume is plotted in the inset Fig. 2(b) in terms of the Eulerian strain [15], $f = \left[(V_0/V)^{2/3} - 1\right]/2$ , and normalized pressure, $F = p/\left[3f(1+2f)^{2.5}\right]$, calculated by fixing $V_0 = 223.61(2)\text{Å}^3$ . A clear anomaly is indeed observed at $f$ = 0.001 ($p$ ~ 0.5 GPa) that can be correlated with the disappearance of the JT distortion and the CO phase. This is consistent with the results obtained from the magnetization measurements at 150 K with different pressures presented in figure 3, where the evolution from the phase separated state to the FM state is evident in the magnetization curves.

For pressure above 0.5 GPa, the linear compressibilities along the *a* and *b* axes, $k_a$ = 0.0021(2) $GPa^{-1}$ and $k_b$=0.0022(3) $GPa^{-1}$ are slightly larger than those corresponding to the $c$ axes, $k_c$=0.0016(1) $GPa^{-1}$, in good agreement with those reported for other manganite systems [16 17 18].

Another interesting feature is the possibility of assisting the magnetic transition with hydrostatic pressure. In figure 3, we also present magnetization as function of magnetic field with different applied pressure at T = 220 K. We can see how the magnetic field necessary to reach an ordered FM state decreases with pressure. To quantify this phenomena, we calculated the entropy change related with the magnetocaloric effect (MCE) using a well know Maxwell relation [2] for different pressures at 220K (figure 4). If we consider the magnetic field necessary to obtain some arbitrary entropy change (i.e. 0.5 J/mol.K) we can see a reduction of the magnetic field of around 2.5 T for each applied GPa. Thereby it is possible to reach the same entropy change with a magnetic field suitable to be produced for general distribution (around 1 T with a pressure of 1.1 GPa). Therefore, it is possible to think of a magnetocaloric refrigerator where the refrigerant material is kept under a fixed pressure during the entire thermodynamic cycle, reaching the same entropy change with a lower applied field.

In summary, we have presented a study of the magnetic and structural properties of the compound $La_{0.625-y}Pr_{0.3}Ca_{0,075}MnO_3$ as a function of the hydrostatic pressure. The presence of the CO state has been observed in the evolution of the lattice parameters due to the Jahn Teller distortion associated with the charge localization. The CO state can be suppressed once it has been formed in the system as the possible applications for barocaloric effect require.

Furthermore, the possibility of assisting the magnetic transition with external pressure has been explored. We observed an important reduction of the magnetic field necessary to induce an entropy change, revealing the possibility of a hybrid baromagnetic effect based systems for refrigeration. It is important to mention that the pressure range needed to magnify the magnetocaloric effect on this manganite system is accessible for industrial applications. It is clear that the presented system is not adequate for room temperature applications. However, we believe that is a good example of the expected behavior of a “baromagnetic material”. More systematic studies are needed in order to reduce the applied pressure, increase the ferromagnetic transition and the MCE to be optimally designed for room temperature applications.

This work has been done with the support of ANPCyT PICT 1327/08, Conicet PIP 00889 and UNSAM SJ10/13. We acknowledge the ESRF and Mohamed Mezouar for providing in-house beamtime at ID27 and J. Jacobs for the preparation of the diamond anvil cell. We are grateful to Leticia Granja and Joaquin Sacanell for fruitful discussions and to Gary Admans for a careful reading of the manuscript. MQ is also member of CIC-CONICET.

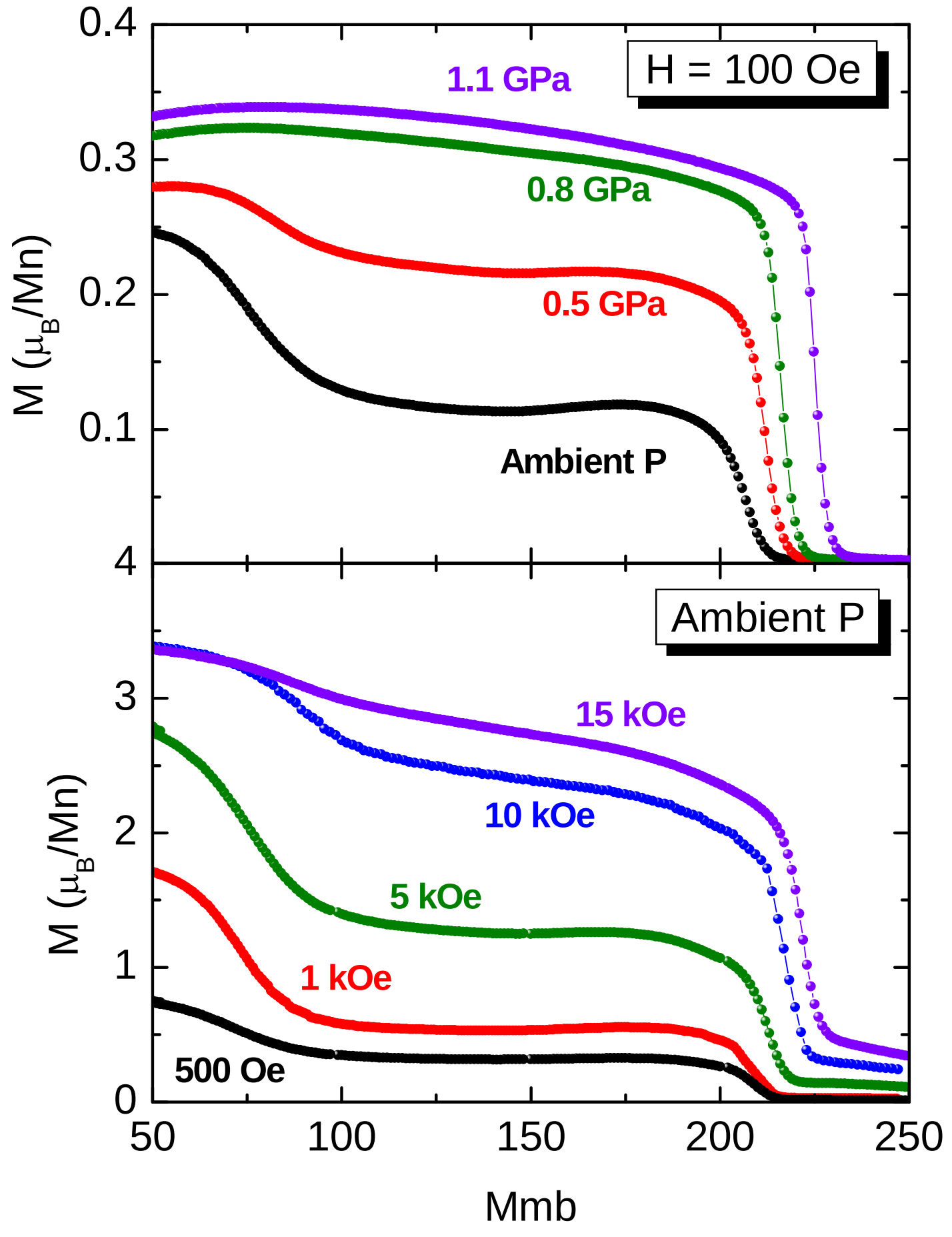

Figure 1: Magnetic moment as function of temperature for different pressures with fixed magnetic field (top panel) and varying the magnetic field at ambient pressure (bottom panel).

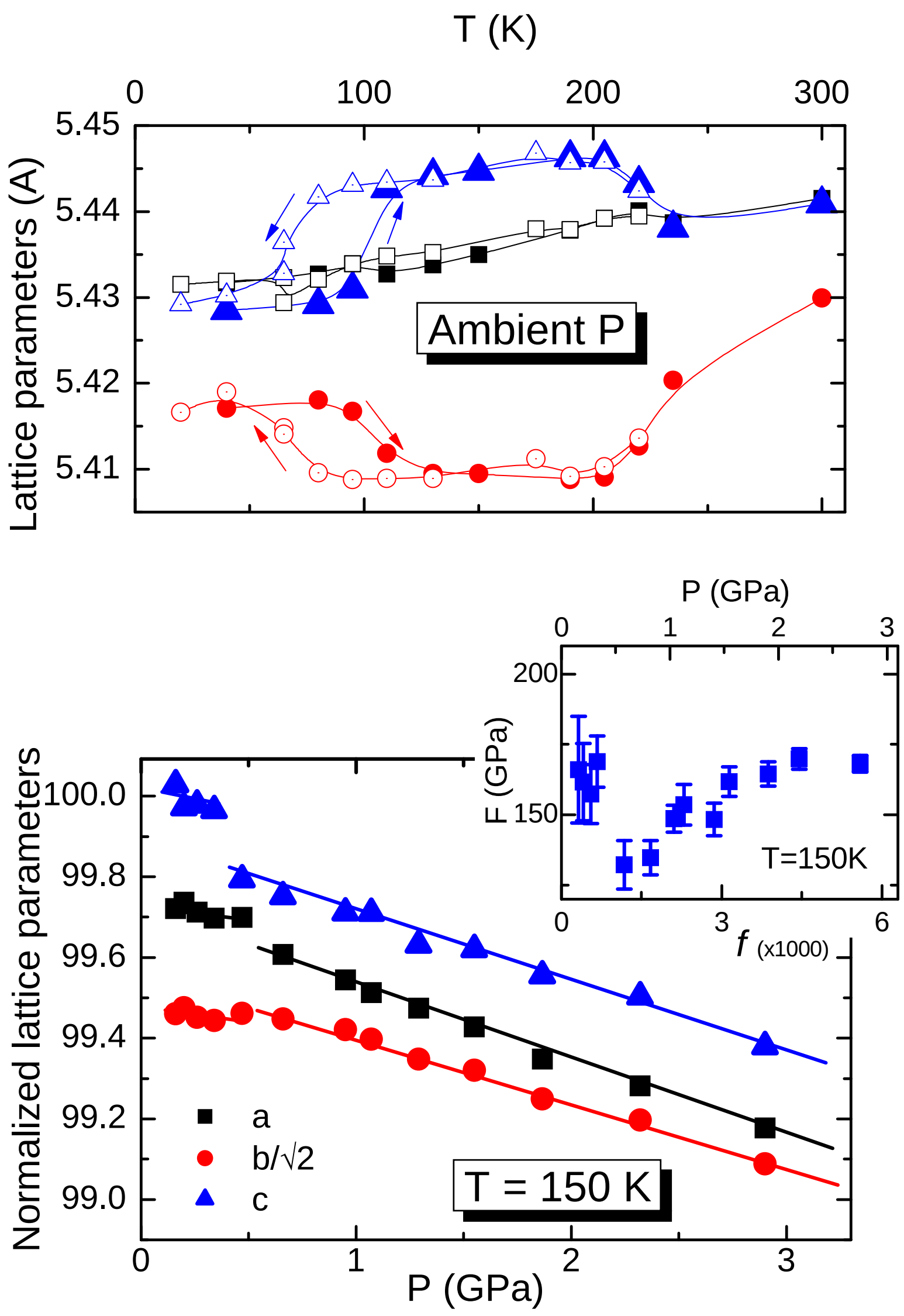

Figure 2: (upper panel) Lattice parameters of LPCMO as function of temperature for cooling (open symbols) and warming cycles (solid symbols). (lower panel) Pressure evolution of the lattice parameters, normalized with the c parameter at ambient pressure. (inset): Eulerian strain, $f$ , versus normalized pressure, $F$ , calculated by fixing $V_0 = 223.61(2)\text{Å}^3$ . The jump observed at ~0.5 GPa can be associated with the suppression of the CO region and the Jahn-Teller distortion.

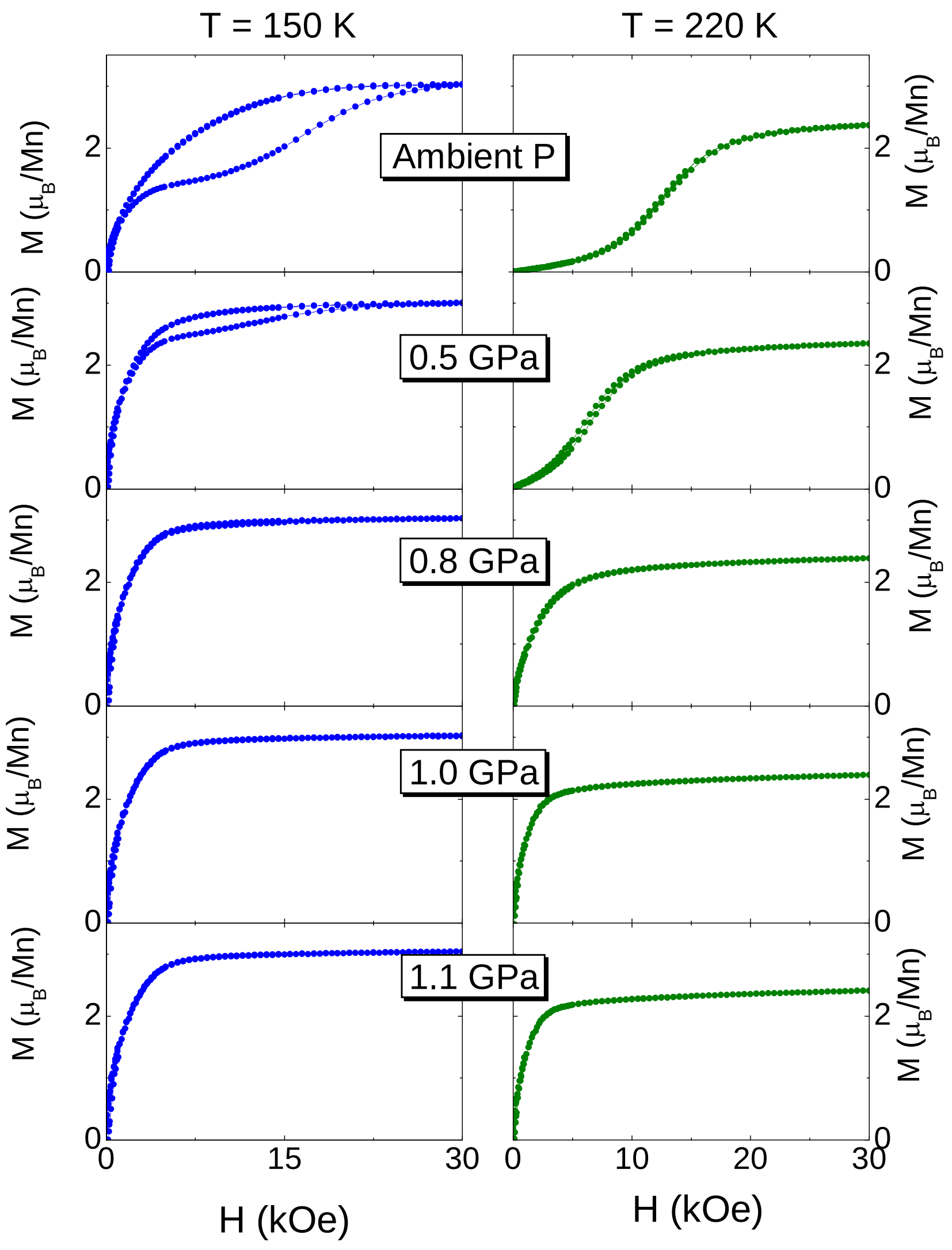

Figure 3:Magnetization as function of the applied magnetic field for different pressures at fixed temperature of 150 K (left) and 220 K (right). It can be clearly seen that above 0.5GPa a transition to a ferromagnetic state is induced.

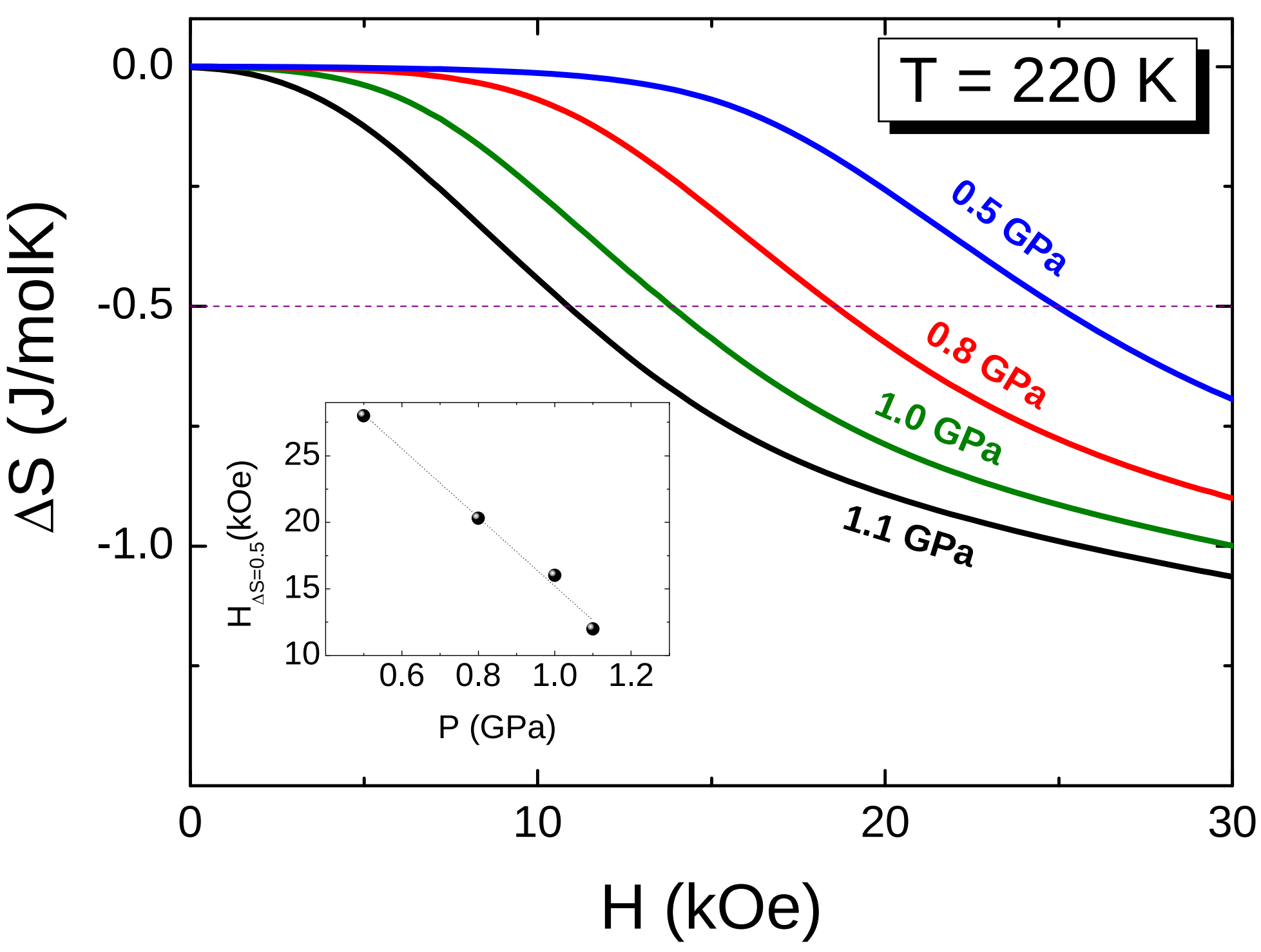

Figure 4: Entropy change as function of magnetic field for different applied pressures at T = 220K using a well know Maxwell relation [2]. In the inset are indicated the magnetic field values necessary to reach an entropy change of 0.5 J/mol.K at the same temperature.

[1] O. Sari, , M. Balli, International Journal of Refrigeration vol. 37, p. 8-15 (2014)
[2] A. M. Tishin and Y. I. Spichkin, The Magnetocaloric Effect and Its Application (IOP, Bristol and Philadelphia, 2003).
[3] N. A. de Oliveira, P. J. Von Ranke, Phys. Rep. 489, 89 – 159 (2010)
[4] Bingfeng Yu , Min Liu , P. W. Egolf , A. Kitanovski, International journal of refrigeration vol. 33, p. 1029-1060 (2010)
[5] A. Kitanovski , P. W. Egolf, International journal of refrigeration vol. 33, p. 449-464 (2010)
[6] K. A. Gschneider, Jr., V. K. Pecharsky, International Journal of Refrigeration vol. 31, p. 945-961, (2008).
[7] K. Alex Müller, F. Fauth, S. Fischer, M. Koch, A. Furrer and P. Lacorre Appl. Phys. Lett. 73, 1056 (1998)
[8] L. Mañosa, D. González-Alónso, A. Planes, E. Bonnot, M. Barrio, J. Tamarit, S. Aksoy and M. Acet, Nature Materials Vol 9, p478 (2010)
[9] N. A. de Oliveira, Appl. Phys. Lett. 90 , 052501 (2007)
[10] A. V. Kartashev, E. A. Mikhaleva, M. V. Gorev, E. V. Bogdanov, A. V. Cherepakhin, K. A. Sablina, N. V. Mikhashonok, I. N. Flerov and N. V. Volkov J. Appl. Phys. 113 , 073901 (2013)
[11] S. Yuce, M. Barrio, B. Emre, E. Stern-Taulats, A. Planes, J. Tamarit, Y. Mudryk, K. A. Gschneidner Jr., V. K. Pecharsky and L. Mañosa Appl. Phys. Lett. 101 , 071906 (2012)
[12] P. Levy, F. Parisi, M. Quintero, L. Granja, J. Curiale, J. Sacanell, G. Leyva and G. Polla, Phys. Rev. B 65, 140401 (2001)
[13] A. C. Larson and R. B. Von Freele, Los Alamos National Laboratory Report, LUAR pp. 86–748 (1994); B. H. Toby, J. Appl. Crystallogr., 34 210 (2001)
[14] M. Uehara, S. Mori, C. H. Chen, S.-W. Cheong, Nature 399 560 (1999)
[15] R. Jeanloz and R. M. Hazen, American Mineralogist, 76, 1765, (1991)
[16] D P Kozlenko, Z Jirák, I N Goncharenko and B N Savenko, J. Phys.: Condens. Matter 16 5883(2004)
[17] C. Meneghini, D. Levy, S. Mobilio, M. Ortolani, M. Nuñez-Reguero, Ashwani Kumar, and D. D. Sarma, Phys. Rev. B 65, 012111 (2001)
[18] D.P. Kozlenko, V.P. Glazkov, Z. Jirák, B.N. Savenko Journal of Magnetism and Magnetic Materials, Volume 267, Issue 1, 120-126 (2003)